\documentclass[traditabstract,longauth,epsf]{aa}
\usepackage{epsfig}
\usepackage{txfonts}

\def\ms{\,m\,s$^{-1}$\,}         %m.s -1
\def\kms{\,km\,s$^{-1}$\,}         %m.s -1
\def\m2s2{\hbox{\,m$^{2}$\,s$^{-2}$}} %m2.s -2
\begin{document}
\title{Transiting exoplanets from the CoRoT space mission\thanks{The CoRoT space mission, launched on December 27, 2006, has been developed and is operated by CNES, with the contribution of Austria, Belgium, Brazil , ESA (RSSD and Science Program), Germany and Spain.}}
\subtitle{XII. CoRoT-12b: a short-period low-density planet transiting a solar analog star}

%%%% COROT EXO COIs 
\author{ 
M. Gillon\inst{1} 
\and A. Hatzes \inst{2} 
\and Sz. Csizmadia \inst{3} 
\and M. Fridlund\inst{4}
\and M. Deleuil\inst{5}
\and S. Aigrain\inst{6} 
\and R. Alonso\inst{7} 
\and M. Auvergne\inst{8} 
\and A. Baglin\inst{8}  
\and P. Barge\inst{5} 
\and S.~I. Barnes\inst{9}
\and A.~S. Bonomo\inst{5} 
\and P. Bord\'e\inst{10} 
\and F. Bouchy\inst{11,12}
\and H. Bruntt\inst {8}
\and J. Cabrera\inst{3,13} 
\and L. Carone\inst{14} 
\and S. Carpano\inst{4}
\and W.~D. Cochran\inst{15}
\and H.~J. Deeg\inst{16} 
\and R. Dvorak\inst{17} 
\and M. Endl\inst{15}
\and A. Erikson\inst{3}
\and S. Ferraz-Mello\inst{18}
\and D. Gandolfi\inst{4}
\and J.~C. Gazzano\inst{5}
\and E. Guenther\inst{2}
\and T. Guillot \inst{19} 
\and M. Havel\inst{19}
\and G. H\'ebrard\inst{12}
\and L. Jorda\inst{5} 
\and A. L\'eger\inst{10} 
\and A. Llebaria\inst{10} 
\and H. Lammer\inst{20} 
\and C. Lovis\inst{7} 
\and M. Mayor\inst{7}
\and T. Mazeh\inst{21} 
\and J. Montalb\'an\inst{1}
\and C. Moutou\inst{5} 
\and A. Ofir\inst{21}
\and M. Ollivier\inst{10} 
\and M. P\"atzold\inst{14} 
\and F. Pepe\inst{7}
\and D. Queloz\inst{7}
\and H. Rauer\inst{3,22} 
\and D. Rouan\inst{5}
\and B. Samuel \inst{10}
\and A. Santerne\inst{5}
\and J. Schneider\inst{13} 
\and B. Tingley\inst{16} 
\and S. Udry\inst{7}
\and J. Weingrill\inst{20}
\and G. Wuchterl\inst{2} 
}

\offprints{michael.gillon@ulg.ac.be}
\institute{
        Universit\'e de Li\`ege, All\'ee du 6 ao\^ut 17, Sart Tilman, Li\`ege 1, Belgium
\and Th\"uringer Landessternwarte, Sternwarte 5, Tautenburg 5, D-07778 Tautenburg, Germany
\and Institute of Planetary Research, German Aerospace Center, Rutherfordstrasse 2, 12489 Berlin, Germany
\and Research and ScientiÞc Support Department, ESTEC/ESA, PO Box 299, 2200 AG Noordwijk, The Netherlands 
\and Laboratoire d'Astrophysique de Marseille, 38 rue Fr\'ed\'eric Joliot-Curie, 13388 Marseille cedex 13, France
\and Department of Physics, Denys Wilkinson Building Keble Road, Oxford, OX1 3RH
\and Observatoire de l'Universit\'e de Gen\`eve, 51 chemin des Maillettes, 1290 Sauverny, Switzerland 
\and LESIA, Observatoire de Paris, Place J. Janssen, 92195 Meudon cedex, France
\and Department of Physics and Astronomy, University of Canterbury, Private Bag 4800, Christchurch 8140, New Zealand
\and Institut d'Astrophysique Spatiale, Universit\'e Paris-Sud 11 \& CNRS (UMR 8617), B\^at. 121, 91405 Orsay, France 
\and Observatoire de Haute Provence, 04670 Saint Michel l'Observatoire, France
\and Institut d'Astrophysique de Paris, UMR7095 CNRS, Universit\'e Pierre \& Marie Curie, 98bis boulevard Arago, 75014 Paris, France
\and LUTH, Observatoire de Paris, CNRS, Universit\'e Paris Diderot, 5 place Jules Janssen, 92195 Meudon, France
\and Rheinisches Institut f\"ur Umweltforschung an der Universit\"at zu K\"oln, Aachener Strasse 209, 50931, Germany 
\and McDonald Observatory, The University of Texas at Austin, Austin, TX 78731, USA
\and Instituto de Astrof\'isica de Canarias, E-38205 La Laguna, Tenerife, Spain 
\and University of Vienna, Institute of Astronomy, T\"urkenschanzstr. 17, A-1180 Vienna, Austria
\and IAG-Universidade de Sao Paulo, Brazil 
\and Universit\'e de Nice-Sophia Antipolis, CNRS UMR 6202, Observatoire de la C\^ote d'Azur, BP 4229, 06304 Nice Cedex 4, France
\and Space Research Institute, Austrian Academy of Science, Schmiedlstr. 6, A-8042 Graz, Austria 
\and School of Physics and Astronomy, Raymond and Beverly Sackler Faculty of Exact Sciences, Tel Aviv University, Tel Aviv, Israel  
\and Center for Astronomy and Astrophysics, TU Berlin, Hardenbergstr. 36, 10623 Berlin, Germany
}

\date{Received date / accepted date}
\authorrunning{M. Gillon et al.}
\titlerunning{CoRoT-12b: a short-period low-density planet transiting a solar analog star}
%%%%%%%%%%%%%%%%%%%%%%%%%%%%%%%%%%%%%%%%%%%%%%%%%%%%
\abstract{We report the discovery by the CoRoT satellite of a new transiting giant planet in a 2.83 days orbit about 
a $V$=15.5 solar analog star ($M_* = 1.08 \pm 0.08$ $M_\odot$,  $R_* = 1.1 \pm 0.1$ $R_\odot$, 
$T_{\rm eff}  = 5675 \pm 80$ K). This new planet, CoRoT-12b, has a mass of $0.92 \pm 0.07$ $M_{Jup}$ 
and a radius of $1.44 \pm 0.13$ $R_{Jup}$. Its low density can be explained by standard models for irradiated planets.

}
%%%%%%%%%%%%%%%%%%%%%%%%%%%%%%%%%%%%%%%%%%%%%%%%%%%%

\maketitle

\section{Introduction}

Because of their special geometric configuration, a wealth of information can be learned 
about transiting extrasolar planets (e.g., Winn 2010), making them very important for our 
understanding of the vast planetary population hosted by our galaxy. They are the only 
exoplanets for which accurate measurements of the mass and radius are available. 
Furthermore, their atmospheric properties can be studied during their transits and occultations 
(e.g., Deming \& Seager 2009).

More than 70 extrasolar planets transiting their parent stars are now known\footnote{See, e.g., 
Jean Schneider's  Extrasolar Planet Encyclopedia at http://exoplanet.eu}, most of which having been
discovered by dedicated photometric surveys. Among these, the CoRoT (Co$nvection$, 
RO$tation$, $and$ $planetary$ T$ransits$) space mission (Baglin et al. 2009) stands out as a 
pionner project. Because of its excellent instrumental capabilities and its low Earth orbit, CoRoT 
can monitor the same fields of view with a very high photometric precision for up to five months. 
This makes possible the detection of planets that would be out of reach for ground-based 
surveys, as demonstrated for instance by its discovery of the first transiting `Super-Earth' 
CoRoT-7b (L\'eger et al. 2009; Queloz et al. 2009), and the first `temperate'  transiting gaseous 
planet CoRoT-9b (Deeg et al. 2010).

We  report here the discovery of a new planet by CoRoT, a `hot Jupiter' called CoRoT-12b that 
transits a $m_V=15.5$ solar analog star. We present  the CoRoT discovery photometry in Sec.~2.  
The follow-up, ground-based observations establishing the planetary nature of CoRoT-12b are 
presented in Sec.~3, while the spectroscopic determination of the parameters of the host star is 
described in Sec.~4. A global Bayesian analysis of the CoRoT and follow-up data is 
presented with its results in Sec.~5. Finally, we discuss the inferred properties of the CoRoT-12 system 
in Sec.~6.

\section{CoRoT photometric observations}

Table 1 presents the ID, coordinates and magnitude of CoRoT-12. This star is located in a field 
near the galactic anti-center direction, in the $Monoceros$ constellation. It was monitored by CoRoT 
from October 24, 2007 to March 3, 2008 (CoRoT run $LRa01$; see Rauer et al. 2009, Carone et al. in prep.). 
 
\begin{table}[h]      
\centering        
\begin{minipage}[!]{7.0cm}  
\renewcommand{\footnoterule}{}     
\begin{tabular}{lcc}       
\hline\hline                 
CoRoT window ID & LRa01 E2 3459 \\
CoRoT ID & 0102671819 \\
UCAC2 ID & 31290403 \\
USNO-A2 ID  & 0825-03015398 \\
USNO-B1 ID & 0887-0101512 \\
2MASS ID   & J06430476-0117471 \\
GSC2.3 ID &  SB3BK006251\\
\\
\multicolumn{2}{l}{Coordinates} \\
\hline            
RA (J2000)  & 06 43 03.76  \\
Dec (J2000) &  -01 17 47.12 \\
\\
\multicolumn{3}{l}{Magnitudes} \\
\hline
\centering
Filter & Mag & Error \\
\hline
B$^a$  & 16.343 & 0.080 \\
V$^a$  & 15.515 & 0.052\\
r'$^a$ & 15.211 & 0.040 \\
i'$^a$ & 14.685 & 0.069 \\
J$^b$  &  14.024 & 0.029 \\
H$^b$  & 13.630 & 0.030 \\
K$^b$  & 13.557 & 0.041  \\
\hline\hline
\vspace{-0.5cm}
\footnotetext[1]{Provided by Exo-Dat (Deleuil et al, 2009).}
\footnotetext[2]{from 2MASS catalog (Skrustkie et al. 2006).}
\end{tabular}
\end{minipage}
\caption{ IDs, coordinates and magnitudes for the star CoRoT-12.}      
\label{startable}      
\end{table}
  
The transits of CoRoT-12b were noticed after 29 days by the so-called `alarm 
mode' pipeline (Surace et al. 2008). The time-sampling  was then changed from 
512s, the nominal value, to 32s. The processed light curve (LC) of CoRoT-12 is shown 
in Fig.~1.  This monochromatic LC  consists of 258~043 photometric measurements for 
a total duration of 131 days. It results from the processing of the raw CoRoT 
measurements by the standard CoRoT pipeline (version 2.1, see Auvergne et al. 2009), 
followed by a further processing (outliers rejection and systematics correction) similar to what
is described by, e.g., Barge et al. (2008) and Alonso et al. (2008). 47 transits of CoRoT-12b 
are present in the LC, 36 of them being found in its oversampled part. Some discontinuities 
are present in the  LC. They were caused by energetic particles hits during the crossings of the 
South-Atlantic Anomaly by the satellite.  A large jump of the measured flux (more than 5\%) 
caused  by the impact of a cosmic ray on the detector can also be noticed in the last part of the
 LC. The processed LC shown in Fig.~1 has an excellent  duty cycle of 91\%.
 
Despite that its CoRoT LC shows some kind of irregular variations with a peak-to-peak amplitude of 
2.3\%, CoRoT-12 appears to be a photometrically quiet star. Except for the transit  signal (see below),
 the discrete Fourier-transform of the LC shows no clear periodicity over the noise 
 level. The rotational period of the star cannot thus be constrained from the CoRoT photometry.

Periodic transit-like signals are clearly visible in the LC, as can be seen in Fig.~1. Initial values 
for the orbital period $P$ and transit epoch $T_0$ were determined by trapezoidal fitting of the transit 
centers, as described by Alonso et al. (2008). The resulting values were $T_0 = 2545398.6305 \pm 
0.0002$ HJD and $P = 2.82805 \pm 0.00005$ days. These values were used to schedule the 
ground-based follow-up observations (see next Section), and also as initial guesses for the global analysis 
presented in Sec. 5.
 
\begin{figure}[h!]
\centering
\includegraphics[width=9cm]{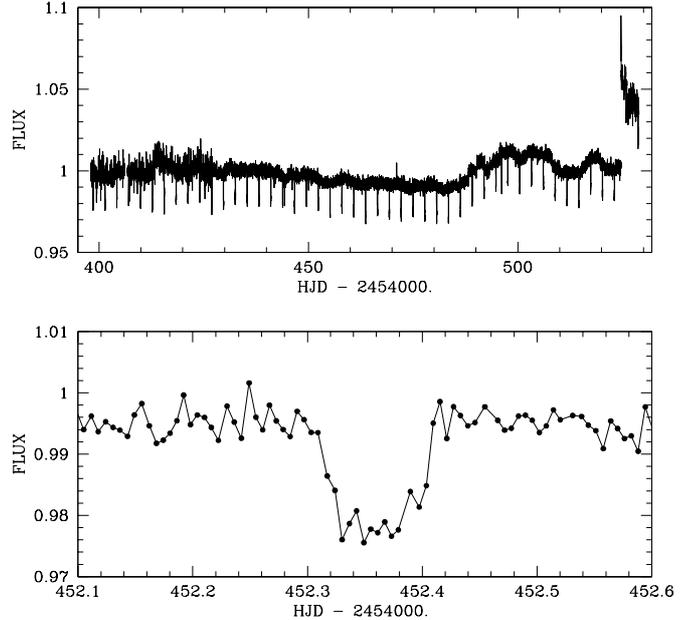}
\caption{$Top$: Normalized CoRoT LC of the star CoRoT-12. The oversampled part of the LC
was binned to the same time bin than its first part for the sake of clarity.
$Bottom$: zoom on a transit of CoRoT-12b.}
\label{fig1}
\end{figure}

\section{Ground-based observations}

The following ground-based observations were performed to establish the planetary nature of 
CoRoT-12b and to better characterize the system.

\subsection{Imaging - contamination} 

CoRoT has a rather poor optical resolution, so performing high-resolution ground-based 
imaging of its fields is important, not only to assess the possibility that the eclipse signals 
detected by CoRoT are due to contaminating eclipsing binaries, but also to estimate the 
dilution of the eclipses measured by CoRoT caused by  contaminating stars (see Deeg 
et al. 2009 for details).

Imaging of the target field was undertaken with the 2.5m INT telescope during pre-launch 
preparations (Deleuil et al., 2009) and with the IAC80 telescope during candidate 
follow-up (Deeg et al. 2009). It found no nearby contaminating star that could be a potential
false alarm source, i.e. that mimiks CoRoT's signal while being an eclipsing binary star (see 
Fig.~2). 

Using the method describe by Deeg et al. (2009), the fraction of contamination in the CoRoT-12 
photometric aperture mask was estimated to be $3.3 \pm 0.5$\%. It is mostly due to a 3.5 mag 
fainter star that is 8.5" SW. This small dilution was taken into account in our analysis presented 
in Sec.~5.

\begin{figure}
\centering
\includegraphics[width=8cm]{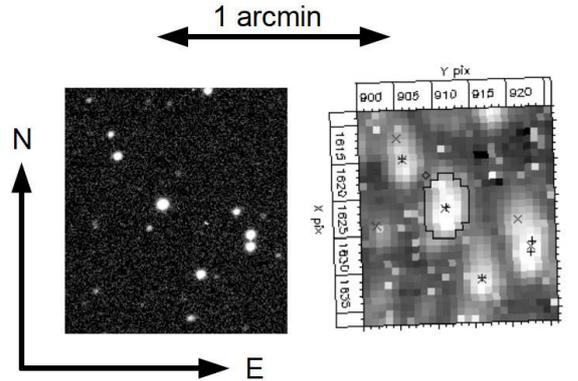}
\caption{The sky area around CoRoT-12 (brightest star near the centre). $Left$:
$R$-filter image with a resolution of 1.3" taken with the INT/WFC. $Right$: Image taken 
by CoRoT, at the same scale and orientation. The jagged outline in its center is the 
photometric aperture mask; indicated are also CoRoT's x and y image coordinates 
and positions of nearby stars from the Exo-Dat (Deleuil et al 2009) database.}
\label{fig2}
\end{figure}

\subsection{Radial velocities - spectroscopy}

Four radial velocity (RV) measurements were obtained with the HARPS spectrograph 
(Pepe et al. 2002b, Mayor et al. 2003) on the 3.6-m telescope at ESO La Silla Observatory 
(Chile), on October 2008 (HARPS program 082.C-0120). These first data were made 
using the high efficiency mode EGGS in order to establish the planetary nature of the 
companion, showing a detectable and low-amplitude radial velocity variation in phase 
with the CoRoT ephemeris, with the shortest exposure time. Ten additionnal 
measurements were recorded with HARPS, from November 27, 2009 to February 05, 
2010 (HARPS program 184.C-0639). These newer data points were acquired using 
the high accuracy mode HAM to increase the precision of the RV measurements compared 
to the about 30 \ms of systematic errors of the high efficiency mode (Moutou et al. 2009), 
and without simultaneous thorium (obj\_AB mode) in order to monitor the Moon background light 
on the second fiber B. Radial  velocities were obtained from the HARPS spectra 
by computing weighted cross-correlation with a numerical G2 mask (Baranne et al. 1996; Pepe et al. 2002a).

Sixteen spectra of CoRoT-12 were also acquired with the HIRES spectrograph on the Keck I 
telescope as part of NASA's key science project in support of the CoRoT mission. 
Differential RVs were computed from these spectra with the $Austral$  code (Endl et al. 2000). 
First, ten spectra were gathered during a transit of CoRoT-12b in January 2009. Unfortunately,
 the used set-up of the slit decker did not allow a proper subtraction of the sky 
 background, leading to RV systematics with an amplitude  of a few dozens \ms, so we 
 decided to reject these data. Six other HIRES RVs were obtained between December 2009 and January 2010.
For these six spectra, the set-up of the slit decker was changed, leading to a proper background subtraction. 

Our HARPS and HIRES measurements are presented in Table 2. An orbital
analysis was performed treating the three sets of RV measurements (HARPS HAM,
HARPS EGGS, and HIRES) as independent data sets with different zero point 
velocities. The orbital solution was made keeping the period and ephemeris 
fixed to the CoRoT values, but allowing the zero point offsets to be fit in a least square
way. Fig.~3 shows the resulting orbital solution which is in phase with the 
CoRoT photometric signal. The resulting eccentricity (0.03 $\pm$ 0.13) was consistent
with zero while the semi-amplitude was 124 $\pm 15$ \ms. Assuming a
solar-mass host star, this semi-amplitude translates into a transiting object with
 a mass of about 0.9 $M_J$. In Sec.~ 5 we present a revised orbit obtained 
 using a global analysis.

The residual RVs were analyzed after removing the orbit to look
for the possible presence of additional companions. No significant variations
were found, but given the sparseness of the measurements we cannot exclude
the presence of additional companions with a good confidence.
 
 The HARPS cross-correlation functions were analyzed using the line-bisector
 technique (Queloz et al. 2000). Fig.~4 shows the correlation between the
bisector and RV  measurements. 
The correlation coeffient of all the RV-bisector measurements, $r$,
has a value of 0.56 with a probability 0.026 that the data is uncorrelated.
Ostensibly this correlation looks to be significant, but we do not believe
that to be the case as this correlation is largely driven by one outlier in the 
HARPS data and another EGGS measurement.
When one examines only the HARPS data the coefficient drops
to $r$ = 0.47 with a probabilty of 0.15 that the data is uncorrelated.
Removing one outlier point  lowers the correlation coefficient to $r$ = 0.32 
with the probability of no correlation being 0.37.

We believe that the modestly high correlation coefficient may be an
artifact of  the bisector error being  more than a factor of two larger
than the RV measurement error and the paucity of measurements.
To test this we generated fake bisector/RV data consisting only
of random noise that was sampled the same way as
the real data. The standard deviations of the fake measurements
were  consistent with the median error
of the RV and bisector measurements. In approximately 40\% of the cases
the correlation coefficient of these random data had correlation coefficients
at least as large as that of the real data. The RV-bisector correlation
coefficient that we measure is consistent with random noise coupled with
sparse sampling. This discards the possibility that the periodic
signal detected in these RVs is caused by a blended eclipsing binary. 
 Taking into account the fact that CoRoT-12 is a solar analog star 
 (see Sect.~4), we interpret thus the eclipses detected in CoRoT 
 photometry as transits of a new giant planet, CoRoT-12b.
 
\begin{table}[h!]
\begin{center}
\begin{tabular}{cccc}
\hline \noalign {\smallskip}
HJD & RV & $\sigma_{RV}$ & Bisector \\  \noalign {\smallskip} 
(days) & (\kms) & (\kms) & (\kms) \\ \noalign {\smallskip} 
\hline \noalign {\smallskip}
HARPS EGGS & & & \\  
\hline \noalign {\smallskip}
2454745.86036 & 12.1740 & 0.0221 & -0.0256  \\ 
2454746.83735 & 11.9341 & 0.0369 & -0.1307 \\ 
2454747.86641 & 12.0904 & 0.0198 & -0.0320 \\ 
2454763.81411 & 11.9856 & 0.0121 & -0.0435  \\ 
\hline \noalign {\smallskip} 
HARPS HAM & & & \\  
\hline \noalign {\smallskip} 
2455163.73528 & 12.1193 & 0.0458 & 0.0429 \\ 
2455165.71941 & 11.9857 & 0.0263 & -0.0310\\  
2455167.72180 & 12.0570 & 0.0342 &  0.0105\\  
2455219.63940 & 12.0051 & 0.0195 & -0.0111\\  
2455220.68849 & 12.2435 & 0.0167 & -0.0188\\  
2455226.66329 & 12.2126 & 0.0168 & 0.0246\\  
2455227.68971 & 12.0348 & 0.0442 & 0.0335\\  
2455229.64150 & 12.2355 & 0.0292 & 0.0259\\  
2455231.68894 & 12.1640 & 0.0240 &  0.0167\\  
2455233.60091 & 11.9993 & 0.0292 & -0.0945\\ 
\hline \noalign {\smallskip}
HIRES  & & & \\  
\hline \noalign {\smallskip}
2455170.99823 & -0.0573 & 0.0286 &  \\  
2455223.00984 &  0.0490 & 0.0144 &   \\ 
2455223.02060 & 0.0466 & 0.0144 &  \\  
2455223.98643 & 0.1633 & 0.0143&  \\  
2455224.93395 & -0.0659 & 0.0175 &   \\ 
2455224.94528 & -0.0979 & 0.0220 &   \\ 
\hline \noalign {\smallskip}
\end{tabular}
\caption{HARPS and HIRES radial velocity measurements for CoRoT-12. The HARPS
RVs are absolute, while the HIRES RVs are differential (measured relative to a stellar 
template). The bisectors were not measured from the HIRES spectra. }
\label{corot-12_rv}
\end{center}
\end{table}

\begin{figure}[h!]
\centering
\includegraphics[width=9cm]{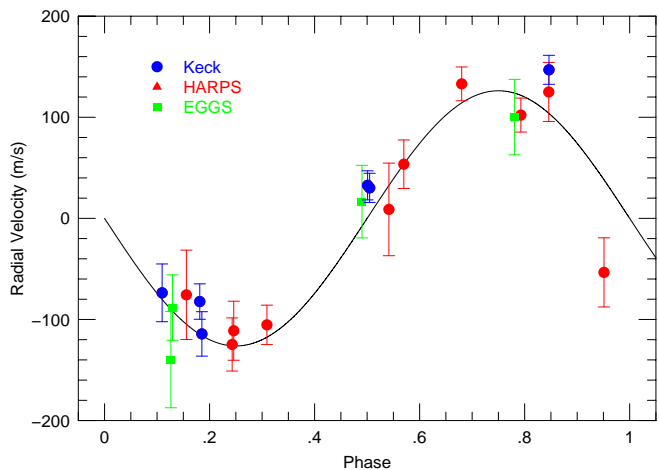}
\caption{HARPS and HIRES RVs phase-folded on the CoRoT ephemeris and overimposed
on the best fit orbital model.}
\label{fig1}
\end{figure}

\begin{figure}[h!]
\centering
\includegraphics[width=9cm]{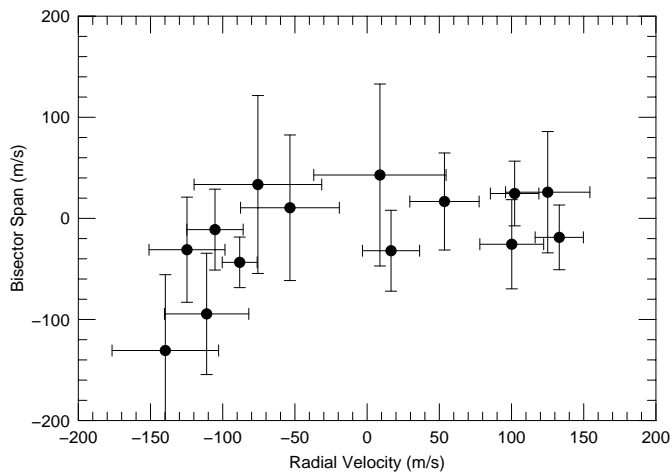}
\caption{Bisector versus RV measured from the HARPS spectra. Errors of twice the RV errors
were adopted for all the bisector measurements.}
\label{fig1}
\end{figure}

\section{Stellar parameters}

Two master spectra were used to determine the atmospheric parameters of the star. The first 
of them was made by co-addition of the seven HARPS HAM spectra which were not
strongly contamined by the Moon background light.  The resulting master spectrum had a 
signal-to-noise ratio (SNR) about 40 in the continuum. The second master spectrum was obtained 
from the co-addition of two Keck spectra and had a SNR about 100 in the continuum.

The methodology used to analyze these two master spectra was mainly based on the semi-automatic package VWA (Bruntt et al. 2002, 2008, 2010), and is thoroughly described by Deleuil et al. (2008) and Bruntt et al. (2010). The derived atmospheric 
parameters and elemental abundances are presented in Table 3.

The Li I line at 670.78 nm was not detected in both master spectra, nor any hint of chromospheric 
activity. From this, the low rotational velocity measured in the spectra, and the low photometric
variability noticed in the CoRoT LC, CoRoT-12 appears thus to be a quiet and slowly rotating 
solar analog star. 

Using T$_{\rm eff}$ and log~$g$ from the VWA spectroscopic analysis, we estimated the absolute 
magnitude M$_V\simeq$~4.75~mag and colour excess ${\rm E}(J-K)\simeq$~0.08~mag from the 
Allen's tables (Cox 2000). We calculated the corresponding interstellar absorption A$_{V}\simeq$~0.46 
(using A$_{V}=(5.82\pm0.1)\times{\rm E}(J-K)$; Cox 2000), to estimate, with the $V$ apparent magnitude, 
the distance of the star to be d~$=1150 \pm 85$ pc.

\begin{table}[h!]
\begin{center}
\begin{tabular}{ccccc}
\hline
\hline \noalign {\smallskip}
$T_{\rm eff}$   &  & & & 5675 $\pm$ 80 K \\  \noalign {\smallskip} 
$\log g$       & & & & 4.52 $\pm$ 0.08 \\ \noalign {\smallskip} 
$\nu_{\rm mic}$   & & & & 0.6 $\pm$ 0.2 \kms \\ \noalign {\smallskip} 
$\nu_{\rm mac}$   & & & & 1.5 $\pm$ 0.3 \kms \\ \noalign {\smallskip} 
$v \sin i$    & & &  & 1.0 $\pm$ 1.0 \kms \\ \noalign {\smallskip} 
$d$    & & &  & 1150 $\pm$ 85 pc \\ \noalign {\smallskip} 
\hline \noalign {\smallskip}
[Fe/H] & & & & 0.16 $\pm$ 0.10 \\ \noalign {\smallskip} 
[Na/H]   & & & & 0.17 $\pm$ 0.06 \\ \noalign {\smallskip} 
[Mg/H]   & & & & 0.13 $\pm$ 0.07 \\ \noalign {\smallskip} 
[Al/H]  & & & & 0.15 $\pm$ 0.10 \\ \noalign {\smallskip} 
[Si/H]   & & & & 0.12 $\pm$ 0.08 \\ \noalign {\smallskip} 
[Ca/H]   & & & & 0.09 $\pm$ 0.10 \\ \noalign {\smallskip} 
[Sc/H]   & & & & 0.22 $\pm$ 0.15 \\ \noalign {\smallskip} 
[Ti/H]   & & & & 0.05 $\pm$ 0.09 \\ \noalign {\smallskip} 
[V/H]    & & & & 0.02 $\pm$ 0.08 \\ \noalign {\smallskip} 
[Cr/H]   & & & & 0.17 $\pm$ 0.09 \\ \noalign {\smallskip} 
[Mn/H]   & & & & 0.20 $\pm$ 0.13 \\ \noalign {\smallskip} 
[Co/H]   & & & & 0.16 $\pm$ 0.14 \\ \noalign {\smallskip} 
[Ni/H]   & & & & 0.21 $\pm$ 0.08 \\ \noalign {\smallskip} 
\hline\\
\end{tabular}
\caption{Stellar parameters and elemental abundances derived for CoRoT-12 from our VWA spectroscopic analysis.}
\label{corot12-params}
\end{center}
\end{table}

\section{Global analysis}

\subsection{Description} 

We performed a thorough global analysis of the CoRoT transit photometry and 
HARPS/HIRES RVs to get the strongest constraints on the system parameters.
First, we cut the parts of the CoRoT LC located within 0.15 days of the  transit mid-times 
deduced from the preliminary transit ephemeris  presented in Sec.~2, getting thus
47 individual transit LCs. Considering their  large number of measurements, we decided to 
stack the measurements of the 36 over-sampled transit LCs per 4, to speed-up our analysis. This
binning did not affect our final precision on the system parameters, as the resulting folded
LC (see Fig.~5) is still well sampled. 

Our analysis was done with the adaptative Markov Chain Monte-Carlo (MCMC) algorithm presented 
 by Gillon et al. (2009; 2010). MCMC is a Bayesian inference method based on stochastic simulations 
 that samples the  posterior probability distributions of adjusted 
parameters for a given model. Our MCMC implementation uses the  Metropolis-Hasting algorithm 
(see, e.g., Carlin \& Louis 2008)  to perform this sampling. Our nominal model was based on  a star and a 
transiting planet on a Keplerian orbit about their center of mass. More specifically, we used a 
classical Keplerian model for the RVs, while we modeled the eclipse photometry with
 the photometric eclipse model of Mandel \& Agol (2002) multiplied  by a baseline model consisting of a
  different fourth-order time polynomial for each of the 47 CoRoT time-series. The coefficients of these baseline models were determined
by least-square minimization at each steps of the Markov chains (see Gillon et al. 2010 for details).

Our analysis was composed of a nominal MCMC run, followed by two other MCMC runs having different specificities that are described below and summarized in Table 4. Each of the MCMC runs 
was composed of five Markov chains of $10^5$ steps, the first 20 \% of each chain being considered as its burn-in phase and discarded. For each run, the convergence of the five Markov chains was checked using the statistical test presented by Gelman and Rubin (1992).

The correlated noise present in the LCs was taken into account as described by
Gillon et al. (2010), i.e., a scaling factor was determined for each LC 
from the standard deviation of the binned and unbinned residuals of a preliminary MCMC analysis,
and it was applied to the error bars (see also Winn et al. 2008). For the RVs, a  `jitter' noise of 5 
\ms  was added quadratically to the error bars, this value being an upper limit for a quiet solar-type 
star like CoRoT-12 (Wright 2005). Practically, this low jitter noise has no impact on the
posterior distributions of the system parameters, as CoRoT-12 is faint and the RV precision is
photon noise/background contamination limited. For the four HARPS measurements obtained 
with the EGGS mode, a systematic error of 30 \ms was also added quadratically to 
the error bars (see Sec.~3.2).

In all three MCMC runs, the following parameters were 
jump parameters\footnote{Jump parameters are the parameters that are randomly perturbed at each step of the MCMC.}:
 the planet/star area ratio $(R_p /R_s )^2$, the transit width (from first to last contact) $W$, 
 the parameter $b' = a \cos{i}/R_\ast$ (which is the transit impact parameter in case of a circular orbit), 
 the orbital period $P$ and time of minimum light $T_0$, the two Lagrangian parameters $e \cos{\omega}$ and $e \sin{\omega}$ 
where $e$ is the orbital eccentricity and $\omega$ is the argument of periastron, and the 
parameter $K_2 = K  \sqrt{1-e^2}   \textrm{ } P^{1/3}$, where $K$ is the RV orbital semi-amplitude
(see Gillon et al. 2009, 2010). We assumed a uniform prior distribution for all these jump parameters.
To take into account the small dilution of the signal due to contaminating stars (see Sec.~3.1), the
jump parameters $(R_p /R_s )^2$ was divided at each step of the MCMC by a number drawn 
from the distribution $N(1.033,0.005^2)$ before being used in the computation of the eclipse model. 

We did not assume a perfectly circular orbit in any of our MCMC runs 
despite that a circular orbit is compatible with the results of our orbital analysis of the RVs (see Sec. 3.2). 
Indeed, most short-period planets could keep a tiny but non-zero eccentricity during a major part of their lifetime (Jackson et al. 2008), so
fixing the eccentricity to zero is not justified by tidal theory and could lead to overoptimistic error bars on the system parameters.

We assumed a quadratic limb-darkening law, and we allowed  the 
quadratic coefficients $u_1$ and $u_2$ to float in our MCMC runs, using as jump parameters not these coefficients 
themselves but the combinations $c_1 = 2 \times u_1 + u_2$  and $c_2 = u_1 - 2 \times u_2$ to minimize the correlation of the obtained uncertainties (Holman et al. 2006).  To obtain a limb-darkening solution consistent with theory, we decided to use  normal prior distributions for $u_1$ 
and $u_2$ based on theoretical values. Sing (2010) presented recently a grid of limb-darkening 
coefficients specially computed for the CoRoT non-standard bandpass and for several limb-darkening laws. We deduced the values $u_1 =0.47 \pm 0.03$ and $u_2 = 0.22 \pm 0.02$ from Sing's grid for the spectroscopic parameters of CoRoT-12 and their errors (Table 3). The corresponding normal distributions $N(0.47,0.03^2)$ and $N(0.22,0.02^2)$ were used as prior distributions for $u_1$ and $u_2$ in our MCMC analysis. 

At each step of the Markov chains, the stellar density deduced from the jump parameters, 
and values for  $T_{ef f}$ and  [Fe/H] drawn from the normal distributions deduced from our 
spectroscopic analysis, were used as input for the stellar mass calibration law deduced by 
Torres et al. (2010) from well-constrained detached binary systems\footnote{The stellar
calibration law presented by Torres et al. is in fact function of   $T_{ef f}$, [Fe/H] and $\log g$.
We modified it to use as input the stellar density instead of the stellar surface gravity (see Anderson
et al. 2010b).}. Using the resulting stellar mass, the physical parameters of the system were then 
deduced from the jump parameters at each MCMC step. To account for the uncertainty on the parameters
of the stellar calibration law, the values for these parameters were randomly drawn at each step
of the Markov chains from the normal distribution presented by Torres et al. (2010).  

In our second MCMC run (labeled $MCMC_2$ in Table 4), we also used as data the parts of the CoRoT LC
located within 0.2 days of the $occultation$ mid-times deduced from the best fit transit ephemeris of our 
nominal $MCMC$ run. The goal of this run was to obtain an upper limit for the depth of the occultation
in the CoRoT photometry. For this run, the occultation depth was thus also a jump parameter.

Finally, we assessed the perfect periodicity of the transits of CoRoT-12b in our third run (labeled $MCMC_3$ in Table 4). 
For this run, a transit timing variation (TTV) was considered as jump parameter for each of the 47 transits. Obviously, the 
orbital period could not be determined unambiguously without any prior on these TTVs, so we assumed a normal prior 
distribution centered on zero for each of them. Practically, we added the following Bayesian penalty to our merit function:
\begin{equation}\label{eq:5}
BP_{\rm timings} = \sum_{i=1,47} \bigg(\frac{TTV_i}{\sigma_{TT_i}} \bigg)^2
 \end{equation}\noindent where $TTV_i$ is the TTV for the $i^{th}$ CoRoT transit, and $\sigma_{TT_i}$ is the error
on its timing estimated by a preliminary individual analysis of this transit. 

\begin{table*}
\centering
\label{tab:params}
\begin{tabular}{lccccccccl}
\hline
& Data & & & Jump parameters & & & Normal prior distributions \\ \noalign {\smallskip}
\hline \noalign {\smallskip}
\hline \noalign {\smallskip}
$MCMC_1$& CoRoT transits & & & $(R_p /R_s )^2$, $W$, $b'$,   & & &$u_1 \sim N(0.47,0.03^2)$\\ \noalign {\smallskip} 
& HARPS (EGGS+HAM) & & &  $P$, $T_0$, $K_2$, $c_1$, $c_2$, & & &$u_2 \sim N(0.22,0.02^2)$ \\ \noalign {\smallskip} 
& HIRES (not transit) & & &  $e \cos{\omega}$, $e \sin{\omega}$ & & &  \\ \noalign {\smallskip}
\hline \noalign {\smallskip}
$MCMC_2$ & idem $MCMC_1$ & & & idem $MCMC_1$&& & $u_1 \sim N(0.47,0.03^2)$ \\ \noalign {\smallskip}
& + CoRoT occultations & & & + occultation depth $dF_2$ &  & & $u_2 \sim N(0.22,0.02^2)$  \\ \noalign {\smallskip}
\hline \noalign {\smallskip}
$MCMC_3$ & idem $MCMC_1$ && & idem $MCMC_1$ & & & $u_1 \sim N(0.47,0.03^2)$\\ \noalign {\smallskip} 
& & && + 47 TTVs   & & & $u_2 \sim N(0.22,0.02^2)$ \\ \noalign {\smallskip}
& & & &  & & & $TTV_{i \in [1:47]} \sim N(0, \sigma_{TT,i}^2)$  \\ \noalign {\smallskip}
\hline \noalign {\smallskip}
\end{tabular}
\caption{Specificities of the three MCMC runs performed during our global analysis. See text for details}
\end{table*}

\subsection{Results}

Table 5 present the CoRoT-12 system parameters and 1-$\sigma$ error limits derived from our nominal MCMC run ($MCMC_1$), and for the two other MCMC runs.

Our MCMC analysis presents CoRoT-12b as an inflated Jupiter-mass planet 
($M_p = 0.92 \pm 0.07$ $M_{Jup}$, $R_p = 1.44 \pm 0.13$ $R_{Jup}$) transiting  a solar analog star 
($M_* = 1.08 \pm 0.08$ $M_\odot$, $R_* = 1.1 \pm 0.1$ $R_\odot$). Using the stellar density 
deduced from our MCMC analysis ($\rho_* = 0.77^{+0.20}_{-0.15}$ $\rho_\odot$) and the effective
temperature and metallicity obtained from spectroscopy (Table 3), a stellar evolution modeling based
on the code CLES (Scuflaire et al. 2008) leaded to a stellar mass of $1.07 \pm 0.10$ $M_\odot$, in 
excellent agreement with our MCMC result, and to a poorly constrained age of $6.3 \pm 3.1$ Gyr.
It is also worth noticing that the two independent values obtained for the stellar surface gravity from our
spectroscopic and  global analysis are in good agreement (1.4 $\sigma$), indicating the good coherence
of our final solution. 

Fig.~5 presents the period-folded CoRoT photometry binned per two minutes time intervals with the best fit transit model superimposed. The standard deviation of the
residuals of this latter LC is 592 ppm, demonstrating the excellent quality of the CoRoT photometry. 

Our results show that  the limb-darkening coefficients $u_1$ and 
$u_2$ are poorly constrained by the CoRoT transit photometry, despite its good precision. Indeed, the posterior distributions of $u_1$ and $u_2$ are close to the 
prior distributions, indicating that the CoRoT data do not constrain these parameters much. 

Our final precisions on the stellar and planetary masses and radii are not excellent (about 7\% on
the masses and about 10\% on the radii), and more observations are required to 
thoroughly characterize  this system. In this context, improving significantly the precision on the stellar density
(about 20\%) is desirable. Such an improvement could be achieved mostly through a better 
characterization of the orbital parameters $e\cos\omega$ and $e\sin\omega$ with more  RV 
measurements (and possibly occultation photometry). Indeed, an new MCMC analysis assuming
 a perfectly circular orbit leads $>2$ times smaller error bars on 
the planet's and star's radii. The characterization of the system would also benefit 
from  an improved determination of the transit parameters with more high-precision transit photometry, if possible acquired
in a redder bandpass (less significant limb-darkening). 

The results of the run $MCMC_2$ show that the occultation of the planet is not detected in the CoRoT data. We can only put an upper limit on its depth (3-$\sigma$ upper limit = 680 ppm). 
 
 As expected,  the errors on $T_0$ and $P$ are significantly larger for the run $MCMC_3$, but the posterior
 distributions obtained for the other parameters agree well with ones of the other MCMC runs. The resulting
 TTVs are shown in Fig.~6. No transit shows a significant timing variation. Still, the resulting TTV series seems to 
 show a correlated structure. Fitting a sinusoidal function in this series leads to a best-fit period of about 24 epochs
 , i.e. of about 68 days. Nevertheless, the resulting false alarm probability is high, about 15\%, indicating that this
 correlated structure is not very significant. Still, it is interesting to notice that, if we assume a rotational period of 68 days
 for the star and $\sin i=1$, and using $R_* = 1.1$ $R_\odot$,  we obtain a value of 1.2 \kms for  $v\sin i$, in excellent
 agreement with the value derived from our spectroscopic analysis (see Table 3). In this context, a possible 
 interpretation of the low-amplitude structure visible in the TTV series is that it is caused by the rotation of 
 the  surface of the star and its influence on the transit barycenters.
   
\begin{figure}[h!]
\centering
\includegraphics[width=9cm]{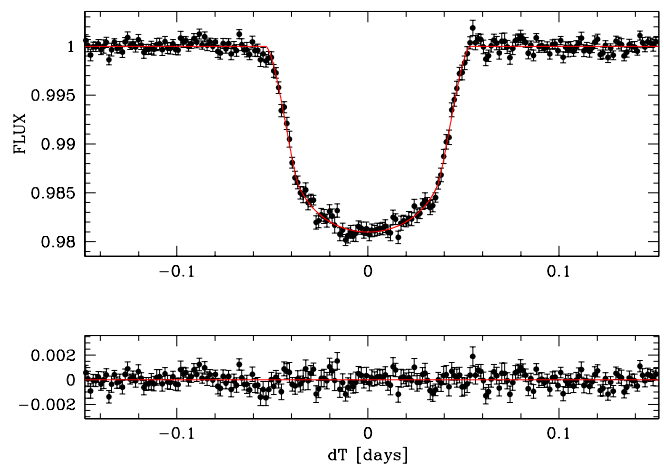}
\caption{$Top$: CoRoT transit photometry period-folded and binned per 2 minutes time intervals, with
the best fit transit model superimposed. $Bottom$: residuals. Their standard deviation is 592 ppm.}
\label{fig2}
\end{figure}

\begin{figure}[h!]
\centering
\includegraphics[width=8cm]{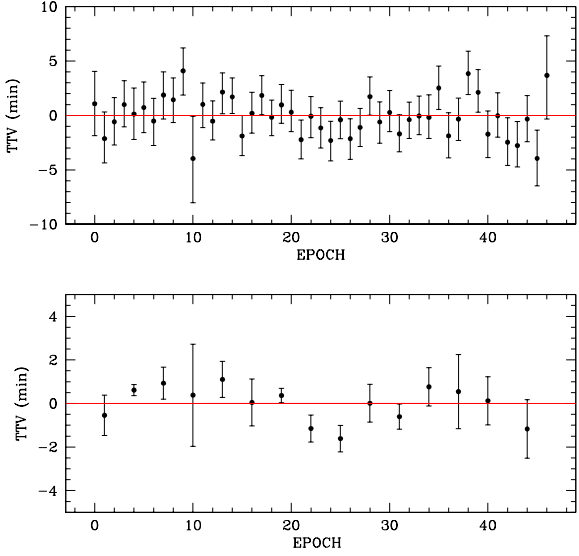}
\caption{$Top$: median value and 1-$\sigma$ limits of the TTV posterior distributions obtained in $MCMC_3$.
$Bottom$: same curve obtained after binning the TTVs per three (error of each bin = error on the mean).}
\label{fig2}
\end{figure}

\begin{table*}[h]
\begin{center}
\begin{tabular}{cccccccccc}
\hline\noalign {\smallskip}
Parameter   & $MCMC_1$ &&&&&&& $MCMC_2$ & $MCMC_3$  \\ \noalign {\smallskip}
\hline \noalign {\smallskip}
$Jump$ $parameters$ & & &&&&&&&  \\ \noalign {\smallskip}
\hline \noalign {\smallskip}
Planet/star area ratio  $ (R_p/R_s)^2 $                & $0.01744^{+0.00039}_{-0.00040}$&&& &&&&$0.01739^{+0.00044}_{-0.00041}$  &$0.01735^{+0.00044}_{-0.00043}$       \\ \noalign {\smallskip} 
$b'=a\cos{i}/R_\ast$ [$R_*$]                                &  $ 0.609^{+ 0.055}_{- 0.057}$               &&&  &&&            & $0.614^{+ 0.060}_{- 0.056}$          & $0.592^{+ 0.040}_{- 0.046}$                 \\ \noalign {\smallskip} 
Transit width  $W$ [d]                                             & $0.10726^{+0.00089}_{-0.00090}$     &&&         &&&     & $0.1071^{+0.0013}_{-0.0011}$       & $0.1071^{+0.0011}_{-0.0013}$              \\ \noalign {\smallskip} 
$T_0-2450000$ [HJD]                                          & $4398.62707 \pm 0.00036$           &&&                 &&&   &  $4398.62704^{+0.00038}_{-0.00036}$  &  $4398.6266^{+0.0013}_{-0.0012}$       \\ \noalign {\smallskip}
Orbital period  $ P$ [d]                                       &$2.828042 \pm 0.000013$         &&&                     &&&    &  $2.828043^{+0.000013}_{-0.000014}$  &  $2.828061^{+0.000052}_{-0.000047}$ \\ \noalign {\smallskip} 
RV $K_2$  [m\,s$^{-1}$\,d$^{1/3}$]                      & $177^{+12}_{-11}$            &&&                                &&& & $176 \pm 11$                                  & $177 \pm 10$                                          \\ \noalign {\smallskip} 
$e\cos{\omega}$                                                     &  $-0.012^{+0.024}_{-0.028}$&&&&&& & $0.000^{+0.020}_{-0.040}$            & $-0.017^{+0.024}_{-0.026}$                   \\ \noalign {\smallskip} 
$e\sin{\omega}$                                                     &  $0.053^{+0.073}_{-0.066}$      &&&   &&&                    & $0.069^{+0.069}_{-0.082}$ & $0.043^{+0.072}_{-0.053}$                     \\ \noalign {\smallskip} 
$c_1$                                                                          & $1.152 \pm 0.056$         &&&                &&&             & $1.153^{+0.054}_{-0.059}$     & $1.146^{+0.058}_{-0.050}$                      \\ \noalign {\smallskip} 
$c_2$                                                                  & $0.028 \pm 0.052$                  &&&                      &&&  & $0.031^{+0.052}_{-0.051}$       & $0.027^{+0.048}_{-0.049}$                       \\ \noalign {\smallskip}                                      
$dF_2$                                                                         &    &&&&&&  &  $0.00009^{+ 0.00022}_{- 0.00009}$     &                                                                  \\ \noalign {\smallskip} 
\hline \noalign {\smallskip}
$Deduced$ $stellar$ $parameters$   &   &   &&&&&&  & \\ \noalign {\smallskip}
\hline \noalign {\smallskip}
$u_1$                                                                             &  $0.466 \pm 0.027$                           &&&         &&&             &  $0.468^{+ 0.026}_{- 0.029}$        &   $0.464^{+ 0.028}_{- 0.025}$         \\ \noalign {\smallskip} 
$u_2$                                                                                   &  $0.219 \pm  0.021$                        &&&   &&&                                & $0.217 \pm 0.020$      &        $0.219^{+0.020}_{-0.019} $      \\ \noalign {\smallskip} 
Density $\rho_* $  [$\rho_\odot $]                                  & $0.77^{+ 0.20}_{- 0.15}$                      &&&            &&&                 & $0.75^{+ 0.20}_{- 0.15}$     & $0.81^{+ 0.18}_{- 0.12}$                  \\ \noalign {\smallskip} 
Surface gravity $\log g_*$ [cgs]                                      & $4.375^{+ 0.065}_{- 0.062}$       &&&       &&&                             & $4.366^{+ 0.066}_{- 0.063}$       & $4.388^{+ 0.055}_{- 0.046}$             \\ \noalign {\smallskip} 
Mass $M_\ast $    [$M_\odot$]&   $1.078^{+0.077}_{-0.072}$ & &&&&&& $1.083^{+0.075}_{-0.074}$       & $1.076^{+0.077}_{-0.071}$                  \\ \noalign {\smallskip} 
Radius  $ R_\ast $   [$R_\odot$]                                 &   $1.116^{+0.096}_{-0.092}$                  &&&    &&&                            & $1.129^{+ 0.097}_{- 0.092} $     & $1.098^{+ 0.072}_{- 0.076} $               \\ \noalign {\smallskip} 
\hline \noalign {\smallskip}
$Deduced$ $planet$ $parameters$   &    &  &&&&&&  &\\ \noalign {\smallskip}
\hline \noalign {\smallskip}
RV $K$ [\ms]                                                                    & $125.5^{+8.0}_{-7.5}$                         &&&    &&&                                   &  $125.4^{+7.4}_{-7.7} $   & $125.5 \pm 7.1 $                                        \\ \noalign {\smallskip} 
$b_{transit}$ [$R_*$]                                                       &   $0.573^{+ 0.027}_{- 0.030}$                &&&         &&&                  & $0.571^{+ 0.031}_{- 0.033}$   & $0.564^{+ 0.033}_{- 0.038}$                       \\ \noalign {\smallskip} 
$b_{occultation}$ [$R_*$]                                                   &  $0.64^{+ 0.10}_{- 0.09} $                     &&&              &&&               &    $0.65^{+ 0.11}_{- 0.09} $         & $0.620^{+ 0.071}_{- 0.078} $                      \\ \noalign {\smallskip} 
$T_{occultation}-2450000$ [HJD]                                       & $4400.020^{+0.055}_{-0.052}$                 &&&                 &&&            & $4400.041^{+0.036}_{-0.073}$     & $4400.010^{+0.043}_{-0.048}$                   \\ \noalign {\smallskip} 
Orbital semi-major axis $ a $ [AU]                                     &   $0.04016^{+0.00093}_{-0.00092}$          &&&                        &&&         & $0.04022^{+0.00091}_{-0.00093}$  &    $0.04013^{+0.00094}_{-0.00090}$         \\ \noalign {\smallskip} 
Orbital inclination $i $ [deg]                                                    & $85.48^{+0.72}_{-0.77}$            &&&                                       &&& & $85.39^{+0.72 }_{-0.84}$  & $85.67^{+0.59 }_{-0.51}$                            \\ \noalign {\smallskip} 
Orbital eccentricity $ e $                                                      &  $0.070^{+0.063}_{-0.042}$                 &&&                                 &&&& $0.083^{+0.062}_{-0.047}$   & $0.059^{+0.061}_{-0.031}$                         \\ \noalign {\smallskip}
Argument of periastron  $ \omega $ [deg]  &  $105^{+90}_{-27}$      &&&     &&&  & $87^{+33}_{-88}$ & $113^{+92}_{-26}$                                       \\ \noalign {\smallskip} 
Equilibrium temperature $T_{eq}$ [K]$ $$^a$                & $1442 \pm 58$            &&&                &&&                                & $1449^{+60}_{-58} $       & $1431 \pm 47$                                            \\ \noalign  {\smallskip} 
Density  $ \rho_p$ [$\rho_{Jup}$]                                          &  $0.309^{+0.097}_{-0.071}$ &&&         &&&                                  & $0.298^{+0.093}_{-0.069}$      & $0.327^{+0.082}_{-0.058}$                           \\ \noalign {\smallskip} 
Surface gravity $\log g_p$ [cgs]                                         &  $3.043^{+0.082}_{-0.080}$           &&&   &&&                                    & $3.031^{+0.083}_{-0.077}$ & $3.060^{+0.065}_{-0.063}$                         \\ \noalign  {\smallskip} 
Mass  $ M_p$ [$M_{Jup}$]                                               &  $0.917^{+0.070}_{-0.065}$            &&& &&&                                   & $0.916^{+0.068}_{-0.064}$      & $0.915^{+0.068}_{-0.064}$                    \\ \noalign {\smallskip} 
Radius  $ R_p $ [$R_{Jup}$]   &  $1.44 \pm 0.13 $       &&&  &&&   & $1.45^{+ 0.13}_{- 0.12} $   &  $1.41^{+ 0.10}_{- 0.09} $                    \\ \noalign {\smallskip} 
\hline \noalign {\smallskip}
\end{tabular}
\caption{Median and 1-$\sigma$ limits of the posterior distributions obtained for the CoRoT-12 system derived from our three MCMC runs (see Table 4). $MCMC_1$ is our nominal analysis (see text for details). $^a$Assuming $A$=0 and $F$=1. }
\end{center}
\end{table*}

\section{Discussion}

\begin{figure}[h!]
\centering
\includegraphics[width=9cm]{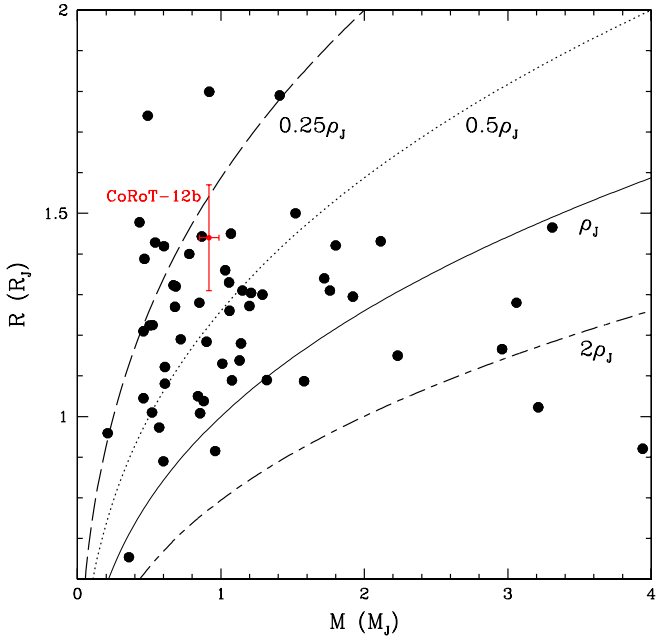}
\caption{Position of CoRoT-12b (in red) among the other transiting planets (black circles, values from http://exoplanet.eu)  in a  mass-radius diagram. The error bars are shown only for CoRoT-12b (C12), WASP-17b (W17), TrES-4b (T4), and WASP-12b (W12)  for the sake of clarity. }
\label{figdens}
\end{figure}

The position of CoRoT-12b in a planetary mass-radius diagram is shown in Fig.~7. While being denser than the 
extremely inflated planets WASP-17b (Anderson et al. 2010a), TrES-4b (Mandushev et al. 2007) and WASP-12b (Hebb et al. 2009), 
CoRoT-12b appears to be a very low-density `hot Jupiter'. Using the hypothesis that the planet is a core-less gazeous planet 
of solar composition, we used the planetary evolution code CEPAN (Guillot \& Morel, 1995) to assess the ability of 
standard irradiated planet models to explain the low-density of CoRoT-12b. Several models were used: a standard model
with no extra heat source, a model for which the opacities were artificially multiplied by 30, and three models with 
a constant energy deposit (10$^{26}$, 10$^{27}$ and 10$^{28}$ erg.s$^{-1}$) at the planet's center. Our results in terms of planetary size evolutions
are shown in Fig.~8. For recall, we constrain the age of the system to  $6.3 \pm 3.1$ Gyr. Considering this age, the measured size of CoRoT-12b is in good agreement with all four evolution models. At most can we notice that  an extra heat source and/or of larger opacities are favored by the data,
but a more precise radius measurement is needed to conclude. 

In this context, it is worth noticing
 that the precision on the planet's radius is mostly limited by the precision on the orbital eccentricity
  and argument of periastron (see Sec. 5.2). It is thus desirable to obtain more RV measurements of 
  the system. Better constraining the planet's orbital eccentricity would also make possible the assessment of its past tidal evolution and its influence on its energy budget (e.g., Ibgui et al. 2010).

\begin{figure}[h!]
\centering
\includegraphics[width=9cm]{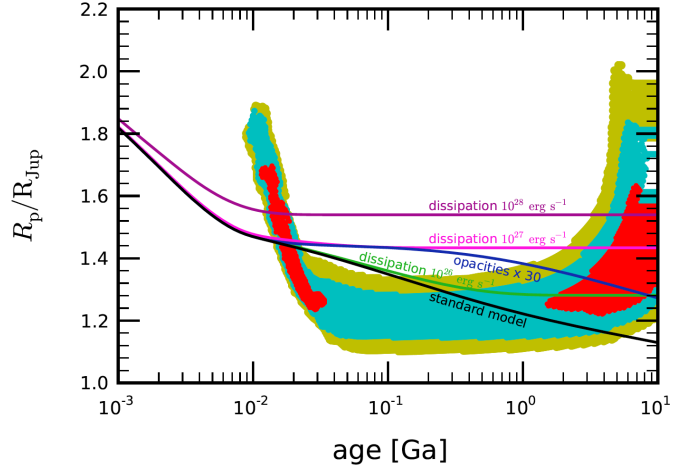}
\caption{ Evolution of the size of CoRoT-12b (in Jupiter units, $R_{Jup}$) 
as a function of age (in billion years), compared to constraints inferred from
 CoRoT photometry, spectroscopy, radial velocimetry and CESAM stellar 
 evolution models. Red, blue and yellow-green dots correspond to 
the planetary radii and ages that result from stellar evolution models matching 
the inferred $\rho_*$-$T_{\rm eff}$ uncertainty ellipse within 1$\sigma$,
2$\sigma$ and 3$\sigma$, respectively. Planetary evolution models for a planet
with a solar-composition envelope and no core are shown as plain lines and are 
labeled as follow~: \textit{standard (black)}: irradiated planet with no extra 
heat source ; \textit{opacities x 30 (blue)}: opacities have been artificially 
multiplied by 30 compared to standard model ; \textit{dissipation}: models in 
which $10^{26}$ ($green$), $10^{27}$ ($pink$), and $10^{28}$ ($purple$) 
${\rm erg\,s^{-1}}$ is deposited at the 
planet's center. These models assume a total mass of 0.92\,$M_{Jup}$ 
and an equilibrium temperature of 1450\,K}
\label{fig2}
\end{figure}

 \begin{acknowledgements}
 The authors wish to thank the staff at ESO La Silla Observatory for their support and for their contribution to the
 success of the HARPS project and operation. 
A part of the data presented herein were obtained at the W.M. Keck Observatory from telescope time allocated to the National 
Aeronautics and Space Administration through the agency's scientific partnership with the California Institute of 
Technology and the University of California. The Observatory was made possible by the generous financial support 
of the W.M. Keck Foundation. The authors wish to recognize and acknowledge the very significant cultural role and 
reverence that the summit of Mauna Kea has always had within the indigenous Hawaiian community. We are most 
fortunate to have the opportunity to conduct observations from this mountain. 
The team at IAC acknowledges support by grant ESP2007-65480-C02-02 of the Spanish Ministerio de Ciencia e 
Innovaci\'on. The building of the input CoRoT/Exoplanet catalog (Exo-dat) was made possible thanks to observations
collected for years at the Isaac Newton Telescope (INT), operated on the island of La Palma by the Isaac Newton group 
in the Spanish Observatorio del Roque de Los Muchachos of the Instituto de Astrophysica de Canarias.
The German CoRoT team (TLS and University of Cologne) acknowledges DLR grants 50OW0204, 50OW0603, and 
50QP07011. The French team wish to thank the "Programme National de Plan\'etologie" (PNP) of CNRS/INSU and the French
National Research Agency (ANR-08-JCJC-0102-01) for their continuous support to our planet search.
The Swiss team acknowledges the ESA PRODEX program and the Swiss National Science Fundation for their
continuous support on CoRoT ground follow-up. M. Gillon acknowledges support from the Belgian Science Policy 
Office in the form of a Return Grant, and thanks B.-O. Demory for useful discussions. Finally, the authors thank the anonymous referee for his valuable comments.
 \end{acknowledgements} 

\bibliographystyle{aa}

\end{document}